\documentclass[useAMS,usenatbib]{mn2e}

\usepackage{comment}

\usepackage{graphicx}
\usepackage{subfigure}

\title[Gravitational nanolensing of quasar dark matter]{Probing planetary mass dark matter in galaxies: gravitational nanolensing of multiply imaged quasars}
\author[H. Garsden,  N. F. Bate  \& G. F. Lewis]{H. Garsden, N. F. Bate and G. F. Lewis\thanks{E-mail:
hgar7294@uni.sydney.edu.au (HG); nbate@sydney.edu.au (NFB); geraint.lewis@sydney.edu.au (GFL). Research undertaken as part of the Commonwealth Cosmology 
Initiative (CCI: www.thecci.org), an international collaboration 
supported by the Australian Research Council.}\\Sydney Institute for Astronomy, School of Physics, A28, The University of Sydney, NSW, 2006, Australia}
\begin{document}

\date{Draft: 9 Sep 2011}

\pagerange{\pageref{firstpage}--\pageref{lastpage}} \pubyear{2010}

\maketitle

\label{firstpage}

\begin{abstract}
Gravitational microlensing of planetary-mass objects (or ``nanolensing'', as it has been termed) can be used to probe the distribution of mass in a galaxy that is acting as a gravitational lens. Microlensing and nanolensing  light curve fluctuations  are indicative of the mass of the compact objects within the
lens, but the size of the source is important, as large sources will smooth out a light curve.
 Numerical studies have been made in the past that investigate a range of sources sizes and masses in the lens. We extend that work in two ways -- by generating high quality maps with over a billion small objects down to a  mass of $2.5\times 10 ^{-5}$M$_\odot$, and by investigating the temporal properties and observability of the nanolensing events. The system studied is a mock quasar system similar to MG 0414+0534.  We find that if  variability of 0.1 mag in amplitude can be observed,   a source size of $\sim 0.1$ Einstein Radius (ER) would be needed to see the effect of  $2.5\times 10^{-5}$M$_\odot$ masses, and larger, in the microlensing light curve. 
Our investigation into the temporal 
properties of nanolensing events finds that there are two scales of nanolensing that can be observed -- one due to
the crossing of nanolensing caustic bands, the other due to  the crossing of nanolensing caustics themselves. The latter are very small, having crossing times of a few days, and requiring sources of size $\sim$ 0.0001 ER to resolve. For sources of the size of an accretion disk, the nanolensing caustics are slightly smoothed-out, but can be observed on time scales of a few days.
The crossing of caustic bands can be observed on times scales of about 3 months.
\end{abstract}


\begin{keywords}
galaxies: structure -- gravitational lensing: micro -- dark matter -- quasars: individual: MG0414+0534 -- methods: numerical 
\end{keywords}

\section{Introduction}\label{Section:Introduction}
Gravitational lensing occurs when light  is deflected by gravity. The first instance of gravitational lensing in a cosmological context was observed in 1979, where the  quasar Q0957+561 is lensed by by a foreground galaxy, 
producing two magnified, distorted, images \citep{walsh1979}. Since then many other multiply-imaged
lensed quasars have been found \citep[e.g.][]{ vanderriest1983, huchra1985, lawrence1995, myers1999, eigenbrod2006}. 
The properties of the lensed images, for example their number, separation on the sky,  relative brightness, and others, are determined mostly by treating the galaxy as a smooth gravitational lens. 
Galaxies are not smooth, however, 
being comprised of many compact objects in addition to smoothly distributed matter. The objects inside the galaxy  can exert an influence on the  quasar images because the quasar and galaxy are in transverse motion relative to our line of sight. 
As the  quasar moves over time, this  motion shifts the light paths relative to the galaxy components, making the quasar images vary in magnitude over time scales of weeks to  years \citep{shalyapin2001}. 
This phenomenon, called gravitational microlensing,   was first observed in 1985 \citep{huchra1985} and  is believed to
occur in most lensed quasars \citep{witt1995}. 
 It can be used as a probe of the source
quasar  because the quasar's size and shape affects the  microlensing variability \citep[e.g.][]{mortonson2005,bate2008b,morgan2010,garsden2010}. 
For example, when the source is large, the changes in the light curve are not as prominent, since the
larger source   smooths out the variability.

Microlensing can also be used to probe the mass distribution in the  lensing galaxy.
Initial studies of quasar microlensing  indicated that  microlensing  was not significantly dependent, within certain constraints, on the
mass spectrum within the lens \citep[e.g.][]{wambsganss1992,lewis1995b}, in particular how much mass is in compact objects and how much is in smooth matter. However, during investigations of anomalous flux
ratios in the lensed quasar MG 0414+0534 \citep{schechter2002}, evidence against this view was found, and 
numerical modelling presented by \citet{schechter2004} showed that it must be false. 
Beginning with a large number of solar-mass  (1M$_\odot$) stars for the lens galaxy, they replaced a substantial fraction of these
with smooth matter, keeping the total mass the same, and found that the microlensing
variability was enhanced. This also introduced the ``bi-modal'' mass distribution into computational microlensing, where
two very distinct mass components in the lens are modelled \citep[e.g.][]{pooley2011}. 

\citet{lewis2006b}  replaced the smooth matter with many compact objects, all of the same mass, and significantly less than  1M$_\odot$. The
 objects produced small-scale microlensing variability that was not present when 
smooth matter was used, as expected. However,
it was found that the variability produced by the small objects could be smoothed out enough so that they could not
be distinguished from smooth matter, if the source size was over a threshold. It was also found that, if source size was taken into account, the amplitude of these variations was an indicator of the mass of the objects in the lens. Recently, \citet{chen2010}  conducted similar modelling,  expanding the bi-modal distribution by using a \citet{salpeter1955}  mass distribution for the stars and   a  \citet{nfw1997} power-law distribution for the compact objects, considering these as potential dark matter candidates.
 They conclude that their small compact objects 
will add detectable small microlensing events (``nanolensing'' events) to a microlensing light curve of no more than about 0.1 mag
over a time scale smaller than a year. \citet{chen2010} were less interested in source size and how this diminishes the detectability of nanolensing events,
but did  confirm that larger sources would smooth out the light curves and make nanolensing harder to detect. 

 The term ``nanolensing'',  used previously by \citet{walker2003} and \citet{schechter2004}, is gaining in usage and refers to  light deflections on much smaller scales than microlensing, due  to planetary-size objects or possible dark matter objects \citep{zakharov2009}. The work of \citet{walker2003} involved the detection of cosmological planetary masses by the nanolensing of gamma ray bursts \citep{walker2003}, and nanolensing is also referred to in exoplanet searches \citep{zakharov2010}.
We use it here to refer to lensing variability produced by objects in a lens with masses far below that of 1M$_\odot$ stars, 
e.g. the low-mass objects in a bi-modal mass distribution.

This paper expands on  the work of \citet{lewis2006b}, \citet{lewis2008} and \citet{chen2010} in investigating nanolensing events due to small masses, and their interaction with source size. Using bi-modal mass distributions, we use several mass values  down to $2.5\times 10^{-5}$M$_\odot$ for the size of the small objects,
while increasing their number to over a billion -- far more than has  been modelled
in the past. We use a mock lensed quasar system that has been modelled by  \citet{schechter2002} and \citet{chen2010}, 
similar to the lensed quasar MG 0414+0534. MG 0414+0534 has not been used in this and past works due to the high magnification of the source quasar, produced by a very large number of objects in the lens galaxy, which are difficult to deal with in numerical models.  Simple statistics allow us to use the amplitude of nanolensing events to infer small objects in a lens, based on a range of source sizes. We then follow this with an investigation into the duration of the  nanolensing events and the source sizes needed to resolve them --
something has not not been done in past studies. We show that there are two time and source scales involved
in these events and indicate how they may be observed. We discuss
how our investigations provide direction to  further work that can be conducted in this area. The structure of the paper is as follows: In Section \ref{method} we introduce numerical modelling of microlensing, and  the lensing model
and  parameters used for this study, in Section \ref{results} we present the results of nanolensing of bi-modal mass 
distributions with
 different source sizes, including event amplitudes and durations. In Section we \ref{discussion} discusses the results and Section \ref{conclusions} contains our conclusions.

Throughout this paper, a cosmology with $H_0 = 70\rmn{km}\rmn{s}^{-1}\rmn{Mpc}^{-1}$, $\Omega_{m} = 0.3$ and $\Omega_{\Lambda} = 0.7$ is assumed.

\section{Method}
\label{method}

\subsection{Numerical analysis of microlensing}

The properties of an image in a multiply-imaged quasar are  determined mostly by the mass distribution of the lensing galaxy, and  the relative locations and distances of the galaxy  and quasar \citep{schneider1992}. For image modelling, two parameters are
used to specify the  mass in the lensing galaxy at the image positions: the convergence ($\kappa$), and shear ($\gamma$). The convergence specifies the effect of mass close to the light path, and the shear is the effect 
of the overall mass of the galaxy. Within the convergence, a mass spectrum can be chosen for models, we will be using  bi-modal distributions as
 described above, where a small amount of mass is in 1M$_\odot$ stars, and the rest in either small objects, or smooth matter. The distances to the lens and source are subsumed into a  distance unit used within lensing models:
the Einstein Radius ($\eta_0$).  If a 
point source is perfectly in line with a point lens object, usually chosen to be 1M$_\odot$, the source will appear as a ring around the lens. Projected onto the source plane, the ring radius is given by
\begin{equation}
\label{Eq:EinsteinRadius}
\eta_0=\sqrt{\frac{4 G M}{c^2} \frac{D_{os} D_{ls}}{D_{ol}}},
\end{equation}
where $M$ is the mass of the lens and $D_{xy}$ refers to the angular diameter distance between $x$ and $y$; the subscripts $s, l,$ and $o$ representing source, lens, and observer respectively.

\begin{figure*}  
\centering
\includegraphics[scale=0.82]{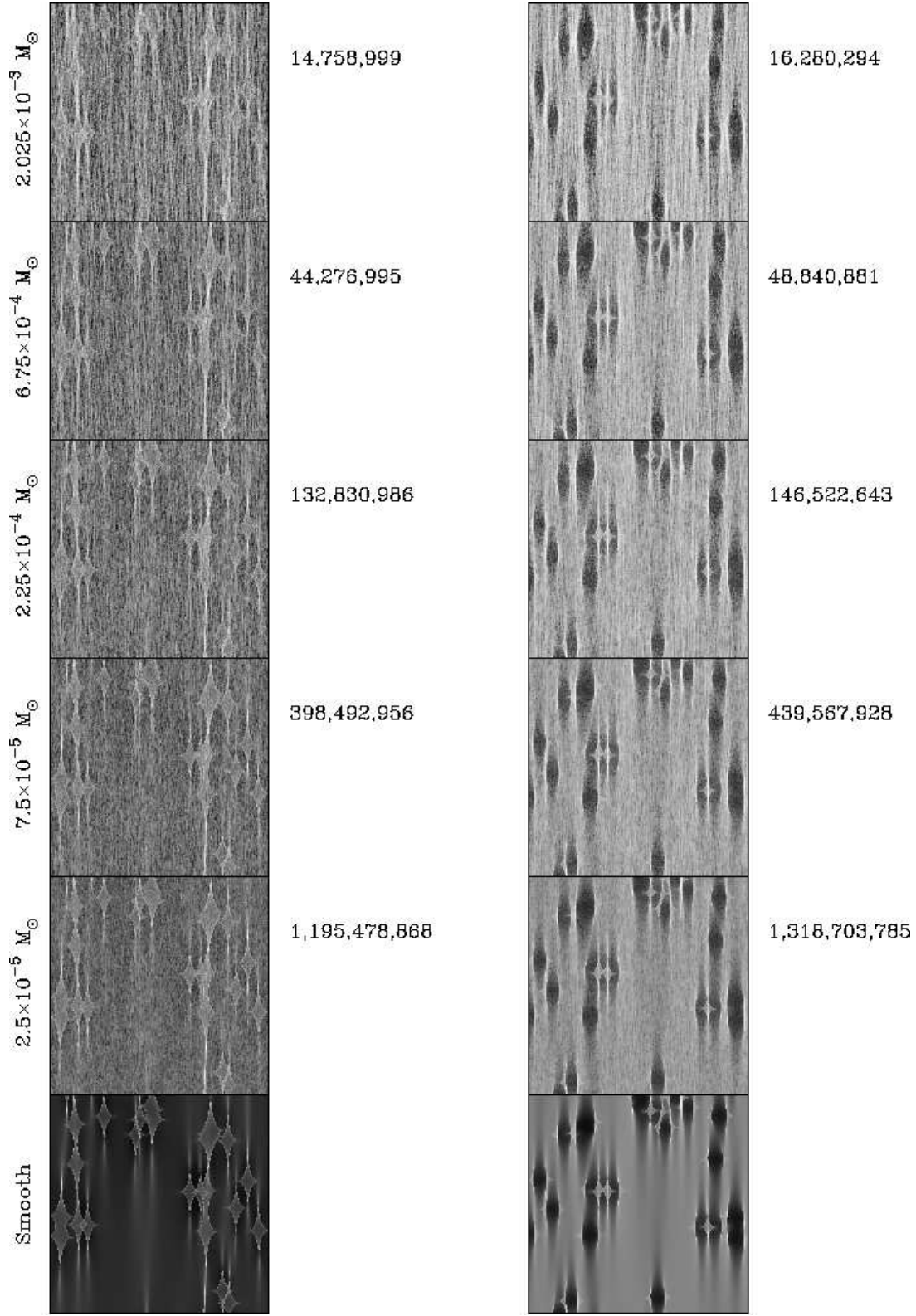}

\caption{The microlensing maps used in this study, all of size 20$\times$20 ER$^2$, with a resolution of 
10000$\times$10000 pixel$^2$. Column 1 contains the maps for image M, generated for a lens
that contains 2\% of the mass in 609 1M$_\odot$ objects, and the rest in either compact objects, or smooth matter.
The mass of the compact objects is indicated on the left side of each map. Column 3 contains the maps for image S, generated for a lens
that contains 2\% of the mass in 672 1M$_\odot$ objects, and the rest in either compact objects, or smooth matter, the compact object mass being the same  as used for the image M map. The mass of the compact matter objects decreases, and the number of compact objects  increases, going down the rows. The second column indicates the number of compact objects for the image M maps, and the 4th column is for the 
image S maps. 
}
\label{maps strip}
\end{figure*}

Microlensing and nanolensing requires the source to change location behind the lens \citep{wyithe2001a,pointdexter2009a}  to produce variation in the source magnification, so a  region of the source plane, where the source may lie, is defined.
Each point in the region has a magnification value, indicating how a point source will be (de)magnified
at that location.
 This is a
``magnification map'' \citep{schmidt2010}, examples of which are seen in Figure \ref{maps strip}. Bright areas in a map indicate locations where the source will 
be  magnified and darker regions are where it will be demagnified, relative to the average for the image being modelled. The regions of light and dark are delineated by lines called ``caustics'', where the magnification is formally infinite \citep{blandford1986}. There will   be different maps  for each image in a lensed quasar, since the mass distribution producing each image is different.

A magnification  map is divided into pixels, which enforces a lower bound on the size of sources that can be studied.   Different source sizes and shapes are studied by creating a pixelized source profile and  convolving this with the map,
producing a map for the microlensing of that source. Light curves can be obtained by taking the magnification along a line across the map, the line corresponding to the path along which a source may travel. For the lensing galaxy,
compact masses and
smooth matter are laid down on a lens plane and rays are fired through them to hit the map. The lens masses are projected onto on a plane  because, in a cosmological situation where the distances are great,  the lensing galaxy can be approximated as flat \citep{saas2003}. Rays are fired in the
inverse direction, i.e. from observer through the lens to source plane, for reasons of computational
ease and efficiency. Because microlensing is produced by a  large number of objects in the lens galaxy, it is difficult to study analytically, and numerical techniques are used; we use the 
inverse
ray-shooting method of \citet{wambsganss1990,wambsganss1999} and \citet{garsden2010}. 
 Numerical approaches have allowed the study of many aspects of quasar microlensing, such as:  the size of accretion disks \citep{chartas2002, bate2008b, blackburne2011}; the  structure of quasar broad line emission regions
\citep{lewis4, keeton, odowd2011}; the structure of microlensed water masers \citep{garsden2011}; chromatic effects in microlensing \citep{1991AJ....102..864W}; the nature of dark matter in lensing galaxies \citep{pooley2009,pooley2011,bate2011};  and the effect of source size on microlensing \citep{bate2007, mortonson2005}, among others.

\begin{table}

 \caption{Parameters for mock lens images}
 \label{parameters}
 \begin{tabular}{@{}ccc}
  \hline
  Image & convergence ($\kappa$) & shear ($\gamma$)  \\
   \hline
   M & 0.475 & 0.425 \\
   S & 0.525 & 0.575  \\
  \hline
 \end{tabular}

 \medskip
 Parameters  for the mock lensed quasar images used in this study. 
 The convergence ($\kappa$) specifies the mass close to a light ray, the shear ($\gamma$) specifies the effect
 of all the mass in the lensing galaxy. They are the same as the M10 and S10 parameters used in \citet{schechter2002}.

\end{table}

\subsection{Mock lens system}

We use  a mock lensed quasar system that has been used in other work \citep{schechter2002,chen2010}. The convergence and shear parameters for each image  are listed in Table \ref{parameters}; these values are based on models of the quadruply-imaged quasar MG 0414+0534. The images in the mock lens system are designated  M and  S; 
image M  is positive-parity,  S is negative parity, the latter meaning the  image is  mirror-symmetric to the source. However, for the purposes of this study,
the internal structure of the images is not relevant, only the magnification of the image relative to the source.
Images M and S  produce different-looking magnification maps and slightly different behaviour in our models, 
as will be seen later.

Recent studies suggest the dark matter fraction in microlensed quasars at the image positions is high. 
\citet{bate2011}  preferred 50$^{+30}_{-40}$\% in MG 0414+0534.  \citet{pooley2009} determined 90\%
for PG 1115+080, and now \citep{pooley2011} suggests $>95$\% for MG 0414+0534 and H1413+117.
For our bi-modal mass distribution, we set two per cent of the mass in the lens  in 1M$_\odot$ objects, 
the other 98\% being in either smooth matter or small compact objects. The small objects are all of the same mass for each map.
The mass of the small objects can be reduced, which means their
number increases to maintain the same total mass in the lens. The 1M$_\odot$ objects stay at  the same locations as the other masses change. 
Following  \citet{chen2010}, we use  z = 0.3 for the lens and z = 2 for the source. The Einstein Radius (ER)
for a 1M$_\odot$ object using these distances is 0.0236 pc.
The magnification maps  cover a region in the source plane of 20 $\times$ 20 ER$^2$ (0.47 $\times$ 0.47 pc$^2$), at a resolution of $10000 \times 10000$ pixel$^2$, 
or  width 0.002 ER (4.7$\times 10^{-5}$ pc)  per  pixel. This also means the smallest source size that can feasibly be studied with such maps is 0.002 ER. 

The maps used in this study are shown in Figure \ref{maps strip}. The first column has maps for image M, the third column for image S.
Note how the large scale caustic structure is consistent within each column, but looks different  between columns. This is
because of the opposite parity of the quasar images \citep{schechter2002}.
The first map for image M was generated from a lens of 609 1M$_\odot$
objects and  14,758,999 objects of $2.025\times 10^{-3}$M$_\odot$, and the second last map  for image M represents 1,195,478,868 objects of $2.5\times 10^{-5}$M$_\odot$. The number of compact objects used to generate each map is listed next to the map in column two. The bottom map was generated from a lens
with 609 1M$_\odot$ objects and the rest of the mass in smooth matter, being the same mass as all the small objects combined. The third column is similar to column one, but for image S, reaching a limit of  1,318,703,785  small objects, with the number of objects listed in column four. To ensure the maps are of high quality with good magnification resolution, we shoot several trillion light rays through the lens, over a million per pixel.

\subsubsection{Computational challenges}

The computational work required to generate and analyse the maps presents considerable  challenges, and it is only in recent years that they have been overcome.
With the advent of supercomputers it is now possible to use large numbers of objects in  the lensing galaxy, and in this work over 1 billion are used. We also fire rays in parallel, using multiple parallel processes. Using the method
of \citet{garsden2010}, the map containing the largest number of masses can be generated in $\sim$ 14 days on a supercomputer  using 16 parallel processes. For the maps with the least number of lens masses, the total time is $\sim$ 24 hours. About 18\% of the compute time is used to generate the lens objects, and the rest  is for firing the rays, with many rays fired in parallel. The need to place the lens information in large data files, of about 150Gb for 1 billion objects, means ideal speedups are not achievable for either  phase of the program, but the situation is improved with the use of 16 Gb of computer memory for caching the files; such memory is only now usable because 64-bit RAM addressing has become available.

\subsection{Source profile and sizes}

Sources in our simulations have a simple 2-D Gaussian profile, with a radius of 3 $\sigma$.
Their radii are based on  estimated sizes of quasar structure.
For example, accretion disks are  of order
$0.002$ pc \citep[][estimated in the I-band from H$\beta$]{mosquera2011}, and broad line regions of order $0.06$ pc \citep[][using the MgII/CIII$\mathrm{]}$ and Civ/Ciii$\mathrm{]}$ lines respectively]{wayth2005, sluse2011}.  For our experiments we use source radii ranging from 0.002  to 1.4 ER, or $4.72\times 10^{-5}$   to 0.033 pc.
Beginning at a size of  0.002 ER, the next size is 0.01 ER, then 0.05 ER, then increasing by 0.05 ER  to 0.4 ER, then in steps of 0.1 ER to 1.0 ER, then in steps of 0.2 ER to 1.4 ER, thus providing more data points at the smaller sizes. This is necessary as will be seen from the figures presented later on.

\subsection{Procedure}
\label{procedure}
We always compare a map generated using stars$+$small objects
 in the lens, to a map using stars$+$smooth matter in the lens, and only maps generated for the same image (M or S). This
means that each map in the first column of Figure \ref{maps strip} will be compared with the bottom one in the column, and
the same for the third column. It can be seen that the maps using small objects become more like the corresponding smooth 
matter map as the size of the small objects gets smaller and their number increases; this
is expected. However, it is also necessary to  consider source size; if these maps are convolved with larger sources 
then they may become similar to the smooth matter case when the small object mass is {\em not} so small.

A simple measure is used to indicate that one map is similar to another. The maps are converted from magnifications to magnitudes. To avoid edge effects, a margin around the map of width equal to the source radius
  of the largest source (1.4 ER) is ignored,
and within the margin the maps are normalized to have an average magnitude of 0. If such a map generated from smooth matter is subtracted from
 a map generated from compact matter the result is an approximately Gaussian distribution of residual magnitude. The 
root-mean-squared (RMS) of the residual values is used as the measure of similarity -- termed the ``difference measure''. It is the average difference between the two maps in units of magnitude.


\begin{figure*} 
\centering
\includegraphics[scale=0.94]{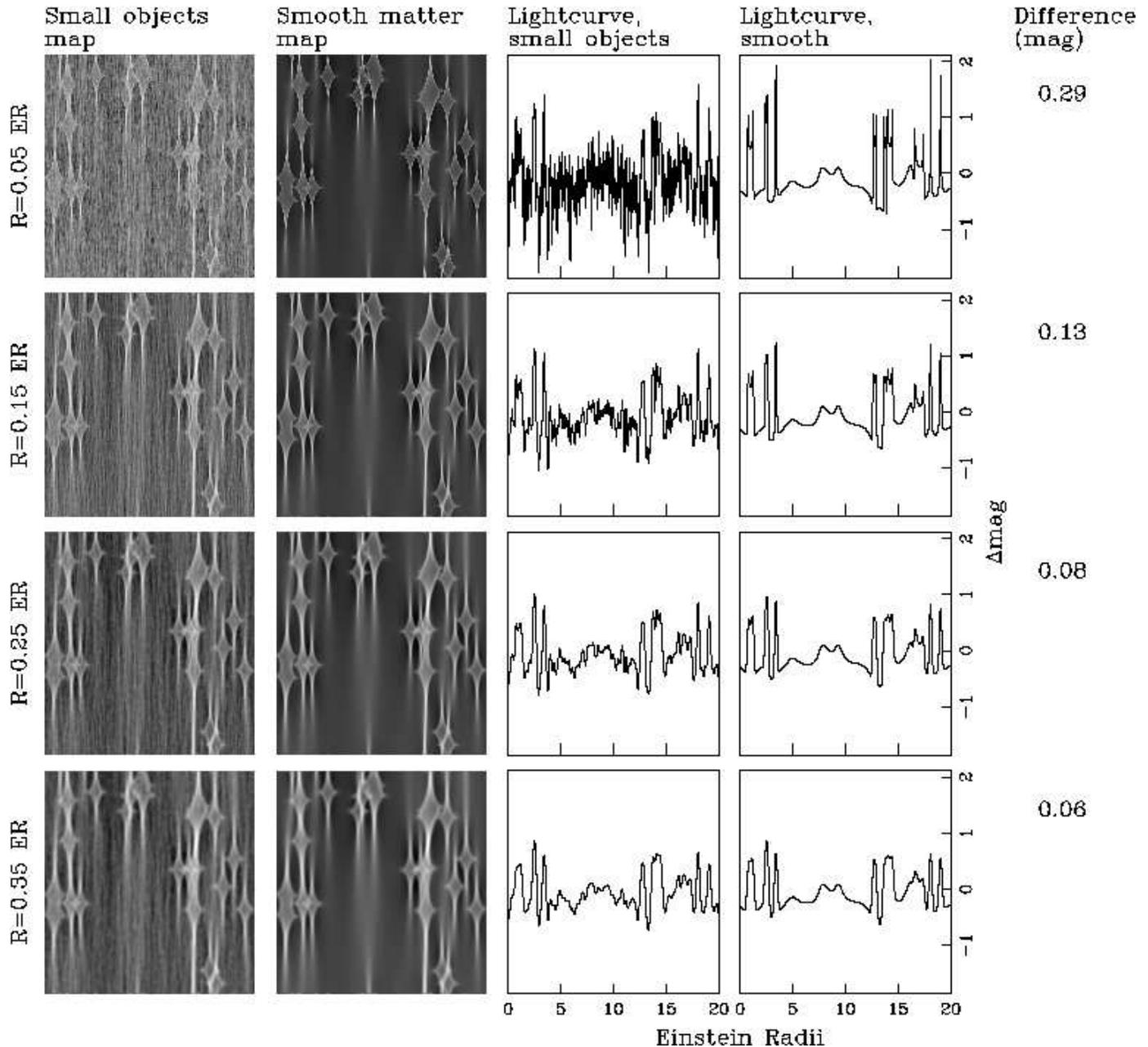}

\caption{Demonstrates how a map generated from compact objects becomes similar to a map generated from smooth matter,
as the size of the microlensed source increases. Column 1  uses the map generated from $2.25\times 10^{-4}$ M$_\odot$ small compact masses for image M in Figure \ref{maps strip}. Column 2 uses the smooth matter map for image M. Column 3 contains a light curve cut horizontally
across the middle of the compact mass map; column 4 contains a light curve from the smooth matter map. Going down the 
rows from top to bottom, both maps are convolved with sources of increasing size, and the size  is indicated on the left side of the compact mass map. Column
5 contains the difference measure  quantifying the difference between the two maps in the row.
}
\label{lightcurves}
\end{figure*}

\section{Results}
\label{results}

\subsection{The effect of increasing source size}

Firstly we demonstrate how two maps, one generated from compact objects and the other from smooth matter, become similar as the 
source size increases. Figure \ref{lightcurves} shows a comparison of such maps; all were  generated from  a star field of 609 1M$_\odot$ masses 
at the same fixed locations, plus either smooth matter or small objects of  $2.25\times 10^{-4}$
M$_\odot$. The first column shows maps for a lens containing these objects, the second column shows maps for a lens
with smooth matter. The 1M$_\odot$ masses  are responsible
for the identical large scale caustics in the maps in each column. The map in the first column shows that between
the large caustics there is small-scale structure, produced by the small objects. The smooth matter map (second column)  does not exhibit
this. The third and fourth columns display a light curve extracted from a horizontal cut across the middle
of the maps in the first and second columns respectively. The light curves make plain the small scale structure in the 
compact matter map, producing a high degree of variability  in the top row that  almost masks the peaks in the light curve from the
large scale caustics, clearly visible in the smooth matter light curve. Going down the row the source size increases, indicated on the left side of the first map. The fifth column is the difference measure between the maps, described previously.

As the source size increases, the compact matter and smooth matter maps and their light curves become similar, and  the difference measure decreases. Both of these things indicate that the increased source size is reducing the small-scale variation in the compact matter map such that it becomes like 
the smooth matter map, and perhaps indistinguishable from it. Our next step is to execute this same procedure for all 
the compact matter maps in Figure \ref{maps strip}.

\subsection{Behaviour of all maps}

\begin{figure*} 
\centering
\subfigure[image M]{\includegraphics[scale=0.48]{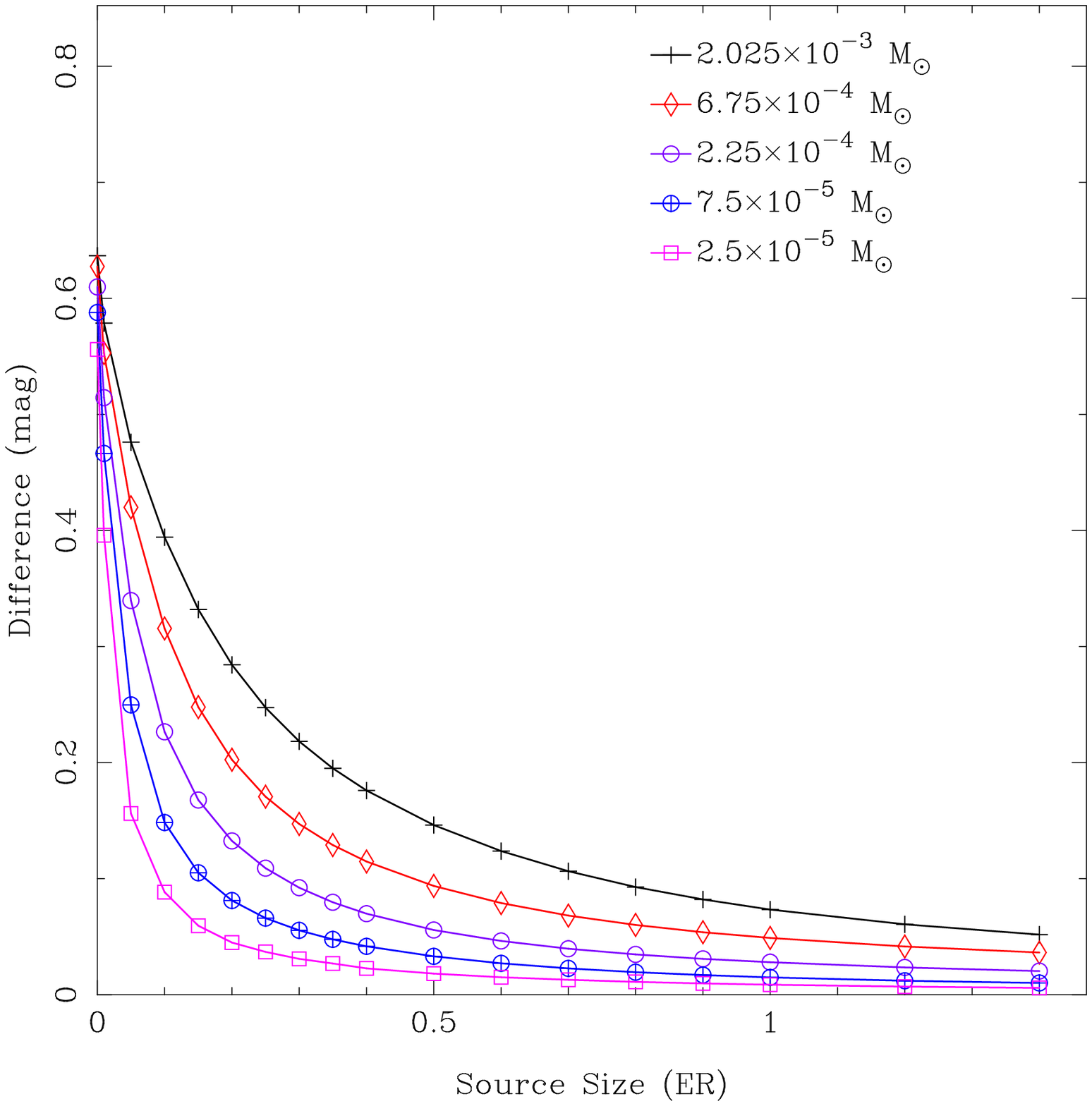}}
\hspace{12pt}
\subfigure[image S]{\includegraphics[scale=0.48]{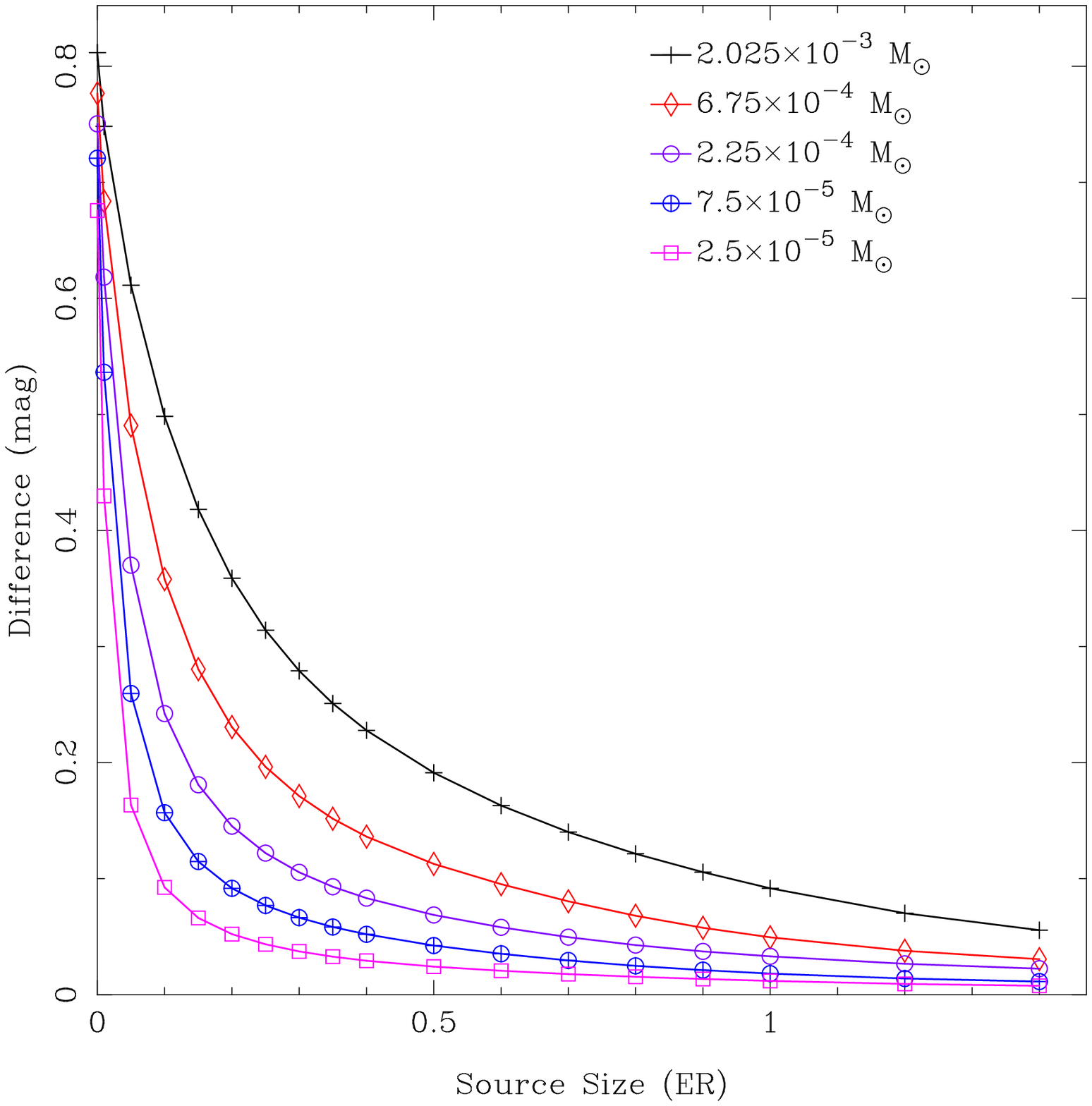}}
\caption{For both images, and for all maps, this is a plot of the difference between the compact matter map and smooth matter map as source size increases. Data for image M is in (a),  image S in (b). In  both graphs there is one curve for each of the five compact matter maps, for each image, in Figure \ref{maps strip}. The top curve corresponds to compact objects of size $2.025\times 10^{-3}$M$_\odot$, the one below for $6.75\times 10^{-4}$M$_\odot$, and so on in decreasing order of mass. The horizontal axis indicates
the size of the source convolved with compact and smooth map for an image, and the vertical axis is an average of the difference  between the compact matter and smooth map.}
\label{graphs}
\end{figure*}

Figure \ref{graphs} shows how the difference measure decreases as source size increases, for all the compact matter
maps. Figure \ref{graphs} (a) is for the maps for image M, and  (b) is for image S. Both plots are similar and show a smooth
decrease in difference (vertical axis) as the source size (horizontal axis) is increased. The difference measure falls faster early on and then flattens out at a value below $\sim 0.05$ for all the maps. The difference starts out smaller for image M,  indicating that the compact object maps are slightly more similar to the smooth matter maps in this image. Maps made with larger masses begin at a higher difference measure, indicating  larger masses produce  higher amplitude variability between the maps. The discriminating power of the difference measure appears to be highest in the middle of the mass
range used.
 Overall, the behaviour is smooth and consistent within each image, and the trend is the same for each image.

\begin{figure*} 
\subfigure[image M]{\includegraphics[scale=0.48]{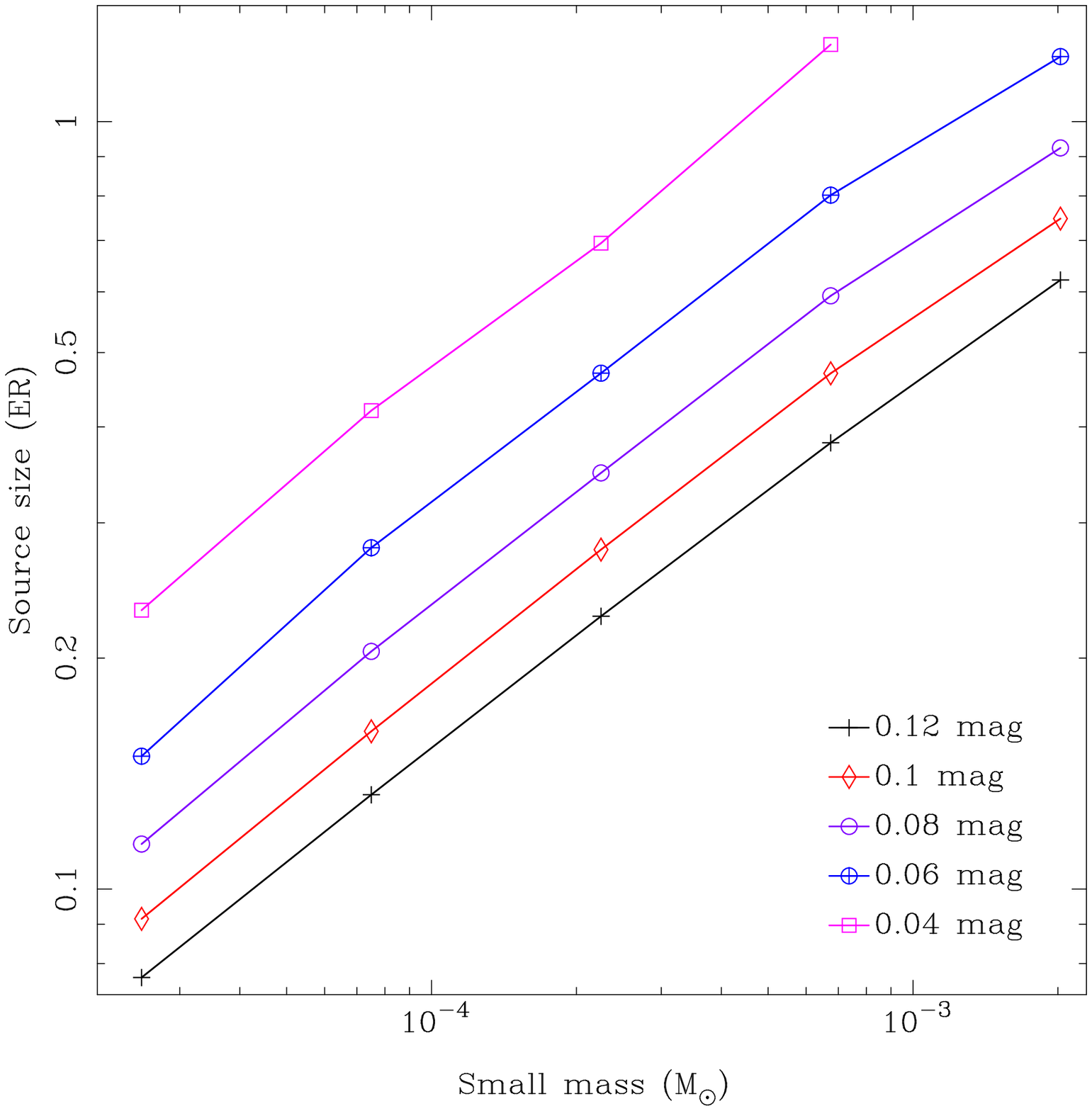}}
\hspace{12pt}
\subfigure[image S]{\includegraphics[scale=0.48]{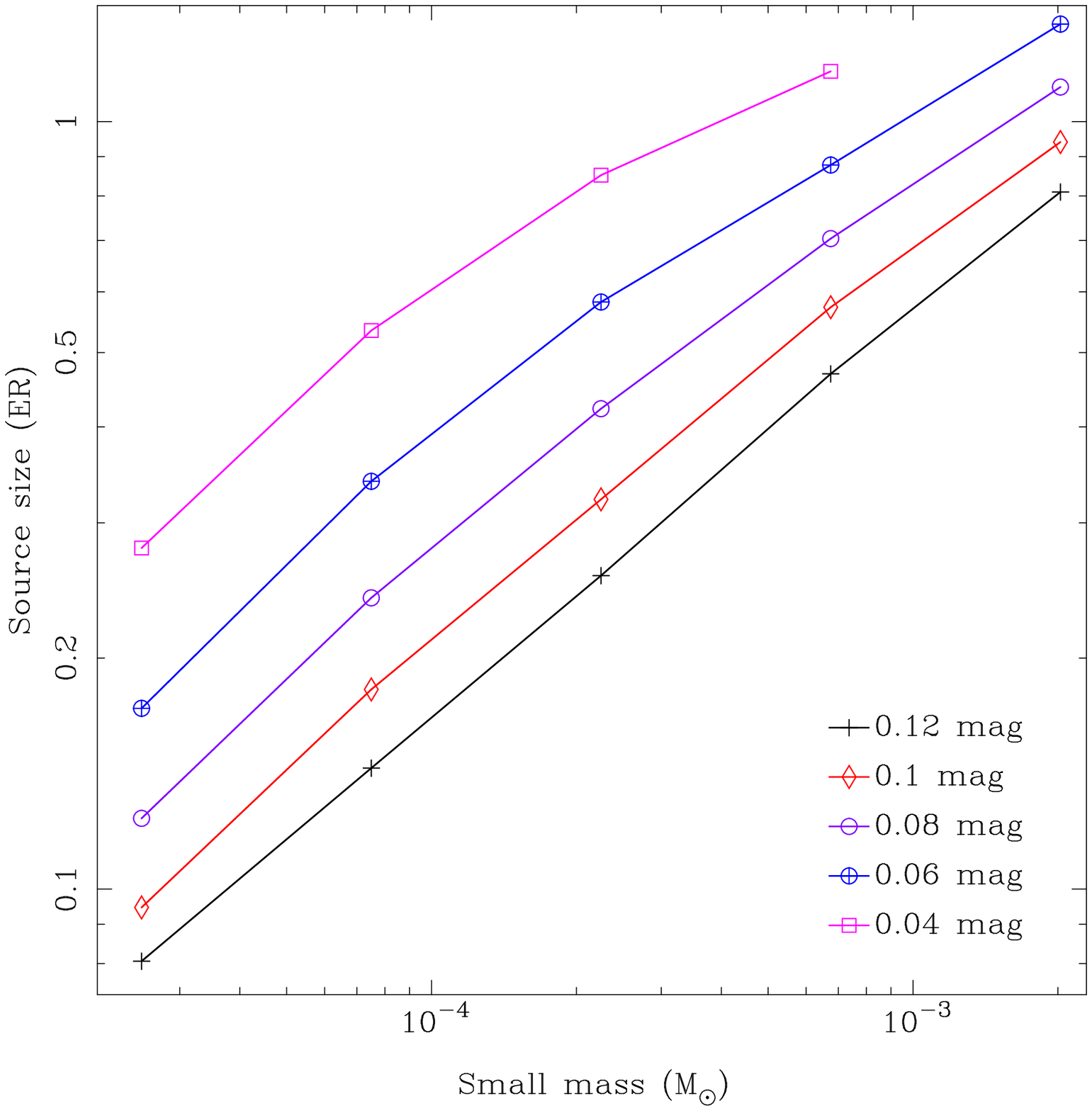}}
\caption{Indicates the maximum size a source can be so that the RMS difference between compact and smooth matter maps 
is greater then 0.12 mag (bottom line) to 0.04 mag (top line), as a function of the mass of the compact objects in the lens. Data for image M is in (a), and image S in (b). The horizontal axis is
 the mass of the compact objects, and the vertical axis is the size of the source being lensed.}
\label{diff_versus_mass}
\end{figure*}

\section{Discussion}
\label{discussion}
As  Figure \ref{lightcurves} has shown, the  difference between the maps is due to nanolensing variability that appears in a light curve as ``wiggles'' that are smaller than microlensing events.  By observing this variability, it is possible to infer that small objects are in the lens, and so distinguish these
objects from smooth matter. Observing  nanolensing requires that the amplitude of nanolensing variability be above the threshold of  observing sensitivity, which we define here as above a certain magnitude threshold. Once nanolensing is found it then depends on its amplitude in relation to the observing threshold as to whether it discriminates between masses for the small objects. If  the threshold is high, then only high amplitude variations can be seen, and from Figure \ref{graphs} this means that only large masses can be inferred in the lens. If the threshold is low, but still only high amplitude variation is seen, it is possible to conclude that smaller masses are \emph{not} in the lens because smaller variations are not seen. Added to this is the size of the source, which reduces the amplitude of nanolensing (and microlensing) variability, so a knowledge of source size is important. We first look at some properties of the map analysis
and then proceed to look at the duration of small-scale changes in light curves, whether these are nanolensing events, and what source sizes are needed to resolve them.

\subsection{Observability of nanolensing events by amplitude}
\label{by amplitude}

It is evident that the simple measure used for the difference between the compact mass and smooth matter maps  exhibits smooth regular behaviour, which we attribute to the physics of the situation, the
high quality of the maps, and the very large number of small objects.  As the behaviour is similar in both images, our difference measure is useful for both positive and negative parity images.  The  measure is an average of the amplitude of nanolensing variation over the whole map,  and hence also an average of light curve deviations, e.g. those seen in Figure \ref{lightcurves}. \citet{chen2010} measured nanolensing from light curve deviations directly,  by finding peaks indicating small-scale changes, \citet{lewis2006b} used  a magnification map analysis similar to ours,  based on fractional differences between  magnification maps.
Using Figure \ref{graphs} we can determine the mass of the small objects in the lens based on the amplitude of nanolensing variation and an assumption of source size. For example, because
all the curves flatten out to below 0.047 mag at a source size of 1.4 ER (0.033 pc in our mock system),  small compact objects will not be distinguishable from each other, or from smooth matter, for a source larger than 1.4 ER, unless 
 the observation threshold is   smaller than 0.047 mag.  

Figure \ref{graphs} can also be used to
correlate nanolensing mass with source size if the amplitude of nanolensing  is  above  a certain threshold.
Figure \ref{diff_versus_mass} shows the largest source size  (vertical axis) that will allow events with amplitudes of   0.12, 0.1. 0.8. 0.6, or 0.04 mag (separate lines) to be observed, for the different masses (horizontal axis). The plots are obtained by  drawing horizontal lines through the graphs in Figure \ref{graphs}. The plot is log-log, which produces a linear trend, as found by \citet{lewis2006b}.
These show that small sources are needed to observe  high amplitude variations, while  a large
source can only produce low magnitude variation.
The source sizes in Figure \ref{diff_versus_mass} range from about 0.07 ER to 1.34 ER (0.0017 to 0.032 pc).  From \citet{chen2010} we adopt a value of 0.1 mag as a threshold difference measure for observing nanolensing. Then, to infer
small  objects of mass $2.5 \times 10^{-5}$M$_\odot$ in the lensing galaxy, the source must be of size 0.09 ER (0.0021 pc) or smaller; for
objects of $2.025\times 10^{-3}$M$_\odot$, a source size of 0.94 ER (0.022 pc) or smaller will do.  The two plots in
Figure \ref{diff_versus_mass}  also show that compact masses are slightly harder to find using 
image M, because the source sizes needed for this are a bit lower overall. This may also be seen in light curves shown later on (Section \ref{obs_by_time}),
where the nanolensing variability appears smaller in image S than that in image M, for the same source.


We may use Figures \ref{graphs} and \ref{diff_versus_mass} to determine the lower limit of the masses that could be discriminated, if the source was  quasar sub-structure such as an accretion disk. For example,  \citet{blackburne2011} estimate the  size of the $\lambda = 590$ nm emission region in  MG 0414+0534 to be $10^{16.23}$ cm, based on microlensing analysis of image flux ratios. This size is 0.23 ER in the mock lens system; using 0.1 mag as the threshold for observing nanolensing in Figure \ref{diff_versus_mass},  variability could be seen for  small masses down to about $1.5\times 10^{-4}$M$_\odot$.
 To see variability caused by  $2.5\times 10^{-5}$M$_\odot$ masses, the source  needs to be 0.09 ER. The $\lambda = 814$ nm emission region in MG 0414+0534 is estimated, from   H$\beta$ flux, to be   $1.9\times 10^{14}$ cm =
$6.2\times 10^{-5}$ pc in MG 0414+0534, or 0.0026 ER in the mock system \citep{mosquera2011}. As this is below 0.09 ER  a source of this size can produce nanolensing  that indicates and discriminates between all the small mass values  we have used. In fact from Figure \ref{graphs}, with a source this small we would observe nanolensing variations of 0.4 mag or better for all the masses.

 \citet{lewis2006b} and \citet{lewis2008} used different  parameters for the lens (e.g. $\kappa$ = 0.2, $\gamma$ = 0.5) and a different criterion for differentiating compact vs smooth, but they suggest  0.2 ER as the source size below which nanolensing  could be distinguished. If we use an observing threshold  of 0.1 mag then 0.2 ER sits nicely within our range, of  0.09 to 0.95 ER, indicating this source size is characteristic of the problem.


The mock lens  used here is similar to that of MG0414+0534, but using   MG 0414+0534 for our analysis  would require 10 billion objects in the lens, if other parameters are held the same. Nevertheless we may make comparisons with MG 0414+0534 by setting the Einstein Radius in the mock lens to the Einstein Radius of MG 0414+0534: 0.0136 pc \citep[z$_L$ = 0.9584, z$_S$ = 2.639;][]{lawrence1995,tonry1998}. In that case, the parsec sizes quoted in the previous discussion  scale down by a factor of 0.58, so that the range in 
Figure \ref{diff_versus_mass} 
becomes 0.0009 to 0.018 pc. \citet{bate2007} find a value of 0.007 pc  for the I-band  emission region in MG 0414+0534, which is within this range, and the Einstein Radius would be 0.52 ER based on 1 ER = 0.0136. For a source of size 0.52 ER, and consulting Figure \ref{diff_versus_mass}, light curve variability due to  masses down to $\sim 6\times 10^{-4}$M$_\odot$ in image S could be observed, at a threshold of 0.1 mag. For an Einstein Radius of 0.0236 pc in the mock lens, no variability over this threshold would be seen
for any mass. We therefore expect that MG 0414+0534 will provide more opportunity for determining small mass size, based on nanolensing amplitude, than the mock lens system.

\citet{chen2010} used a uniform disk for the source profile. A uniform disk with radius R has a half light radius of 0.71 R, a disk with a Gaussian profile and radius of R = 3$\sigma$ has a half-light radius of 0.28 R. Therefore our sources  are equivalent to a uniform disk that is 0.4 the size of the  size stated here.

\subsection{Observability of nanolensing events by time}
\label{obs_by_time}
\begin{figure*}
\centering
\includegraphics[scale=0.85]{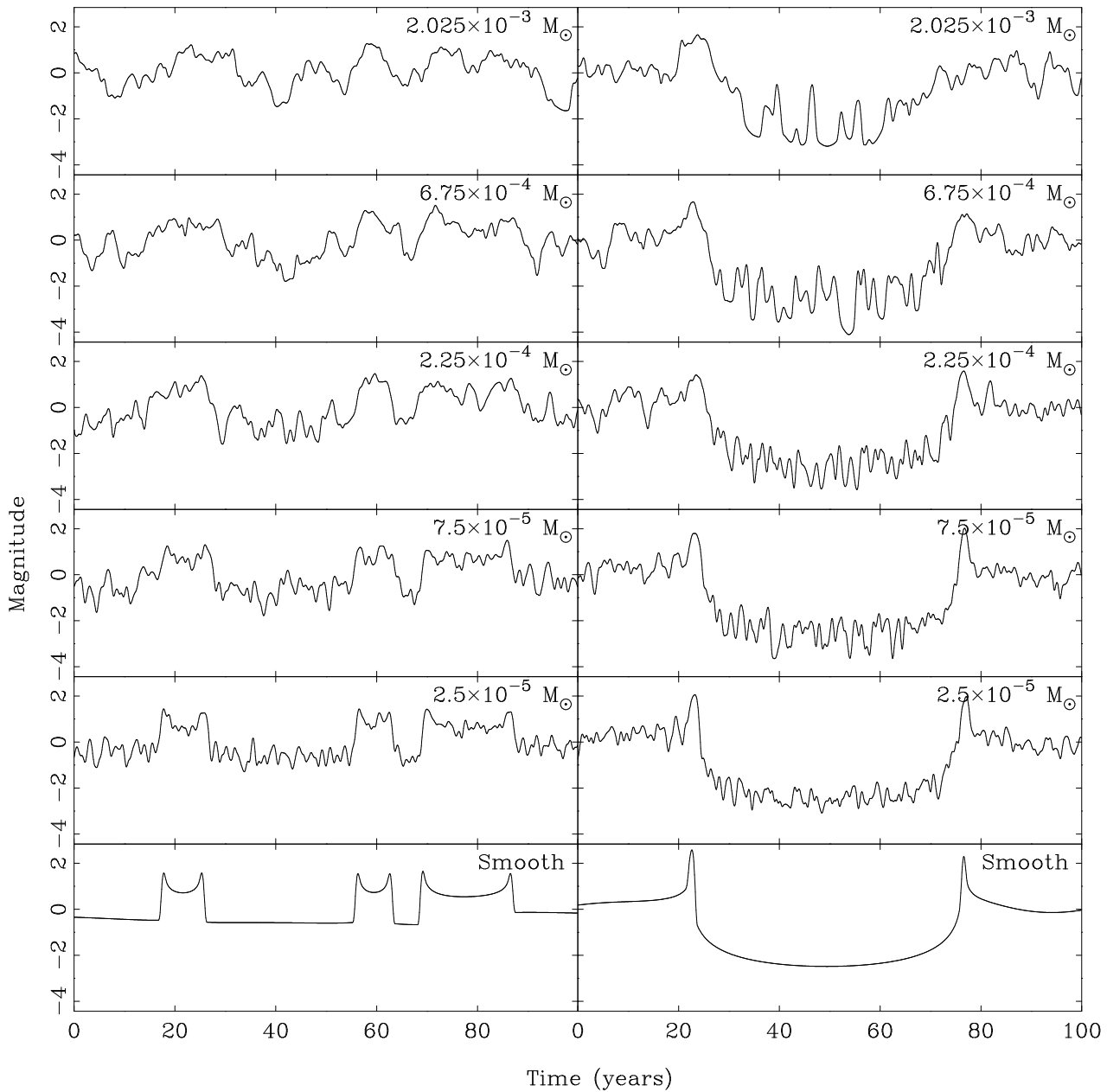}
\caption{Light curves representing a time period of 100 years (2.6 ER, 0.061 pc at 600 km s$^{-1}$), cut horizontally through the middle of the maps in Figure \ref{maps strip}, 
with a source of size 0.02 ER (0.0005 pc). The mass of the small objects used to generate the light curve is indicated in the top right  of each panel, and each
panel corresponds to the map at the same position in Figure \ref{maps strip}.}
\label{wiggles_by_mass}
\end{figure*}

\begin{figure*}
\centering
\includegraphics[scale=0.85]{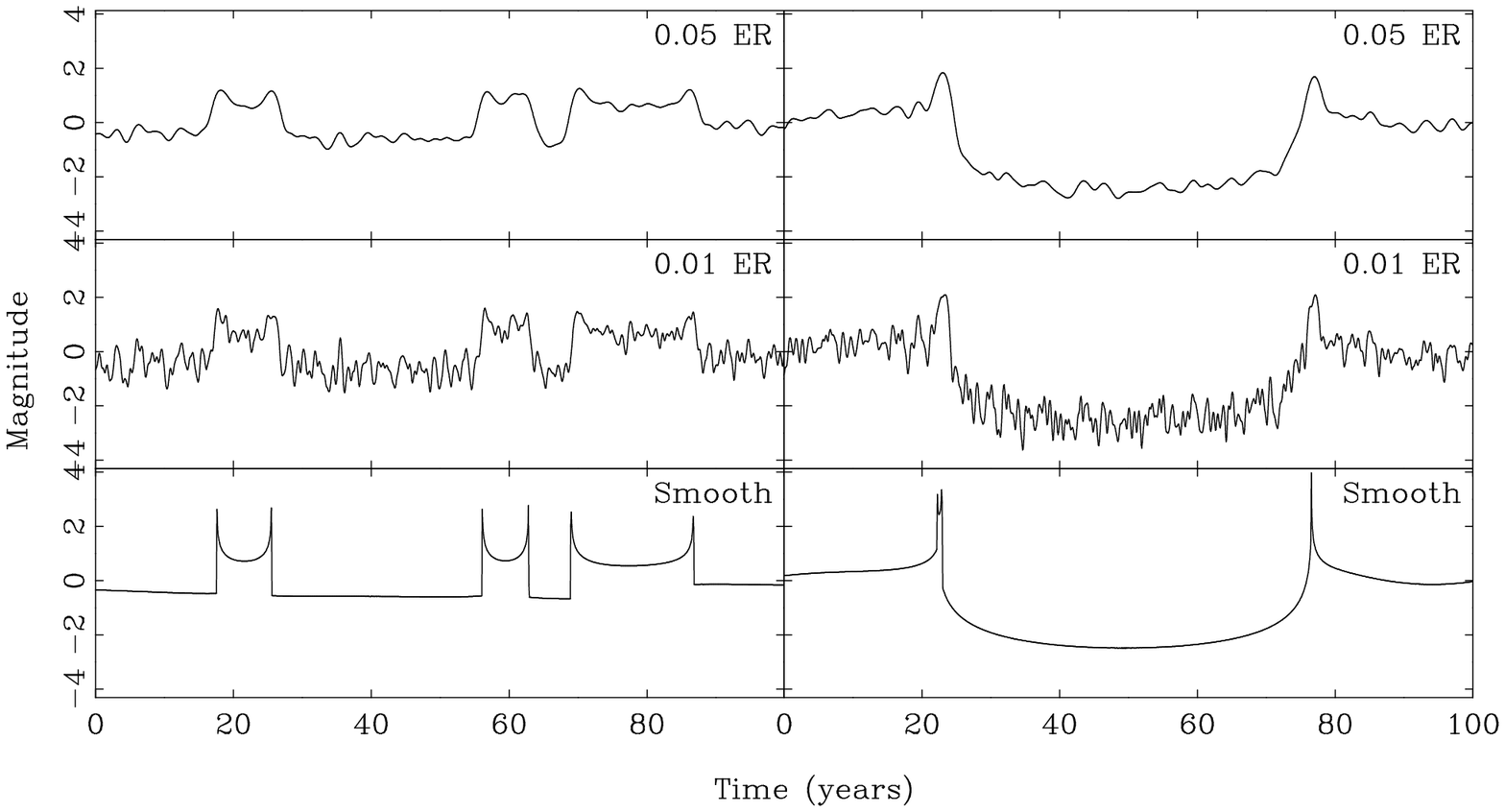}
\caption{Light curves representing a time period of 100 years (2.6 ER, 0.061 pc at 600 km s$^{-1}$), cut horizontally through the middle of the $2.5\times10^{-5}$M$_\odot$ maps in Figure \ref{maps strip}.
The left column uses the $2.5\times10^{-5}$M$_\odot$ map for image M, the right column the same mass map for image S. Each row represents a different
source size, with the size indicated in the top right  of each panel. The bottom row is from the map generated from  smooth matter in place of the  $2.5\times10^{-5}$M$_\odot$ masses, and the source size is the pixel resolution of the maps -- 0.002 ER.}
\label{wiggles_by_size}
\end{figure*}

We  have not yet discussed the temporal characteristics of the nanolensing variation  -- whether it is observable, and how to observe it. In this section we examine these questions, without embarking on a full statistical analysis, which is reserved for future work. We assume a velocity
of the quasar source over the magnification maps of 600 km s$^{-1}$, which has been used in the past for the mock lens system \citep{chen2010}. This gives a time of 38 years  to cross the Einstein Radius of a solar mass star, and 71 days = 0.00012 ER for the Einstein Radius of our smallest mass object ($2.5\times 10^{-5}$M$_\odot$). In a lens where there are many objects and a complex magnification map, such as we have here, we expect events to be seen on timescales smaller than this.

In what follows we have used horizontal light curves relative to all the magnifications maps. The variability is at its greatest, i.e. highest frequency and shortest period,  in that direction, because the shear operates in the vertical direction on the magnification maps, and the caustics are extended in that direction. Sources moving
diagonally across the caustics  will show a  longer time period for nanolensing variations, compared to a horizontally moving source.

 We now turn to an analysis of the nanolensing variations.
Figures \ref{wiggles_by_mass} and \ref{wiggles_by_size} show light curves taken from the image M and image S maps in Figure \ref{maps strip}
in the horizontal direction, over a time period of 100 years. Figure \ref{wiggles_by_mass} uses a fixed source size of 0.02 ER and varies the mass of the small objects, so that each panel represents a different mass, indicated in the top  of each panel. The top row represents the largest mass, $2.025\times 10^{-3}$M$_\odot$,
the next row is $6.75\times 10^{-4}$M$_\odot$, and so on in the same order reading down the page as
in Figure \ref{maps strip}. The bottom row is the smooth matter map with a source size of 0.02 ER.
 Figure \ref{wiggles_by_size} uses the maps that have small objects of  $2.5\times10^{-5}$M$_\odot$,  but varies
the source size, indicated in the top  of each panel. 
The top row represents a source of size of 
0.05 ER, and the next row is 0.01 ER, with the bottom row being the smooth matter map at the pixel resolution of  the maps -- 0.002 ER. In both Figures the left column is image M, the right column is image S.  
The light curves confirm the results of the map analysis, that the amplitude of the nanolensing variability decreases as the small objects or source size decreases. However we have the addition of new temporal information:
as the mass of the small objects or the source size decrease, the frequency of small-scale variation increases.

\begin{figure*} %
\centering
\subfigure
{\includegraphics[scale=0.48]{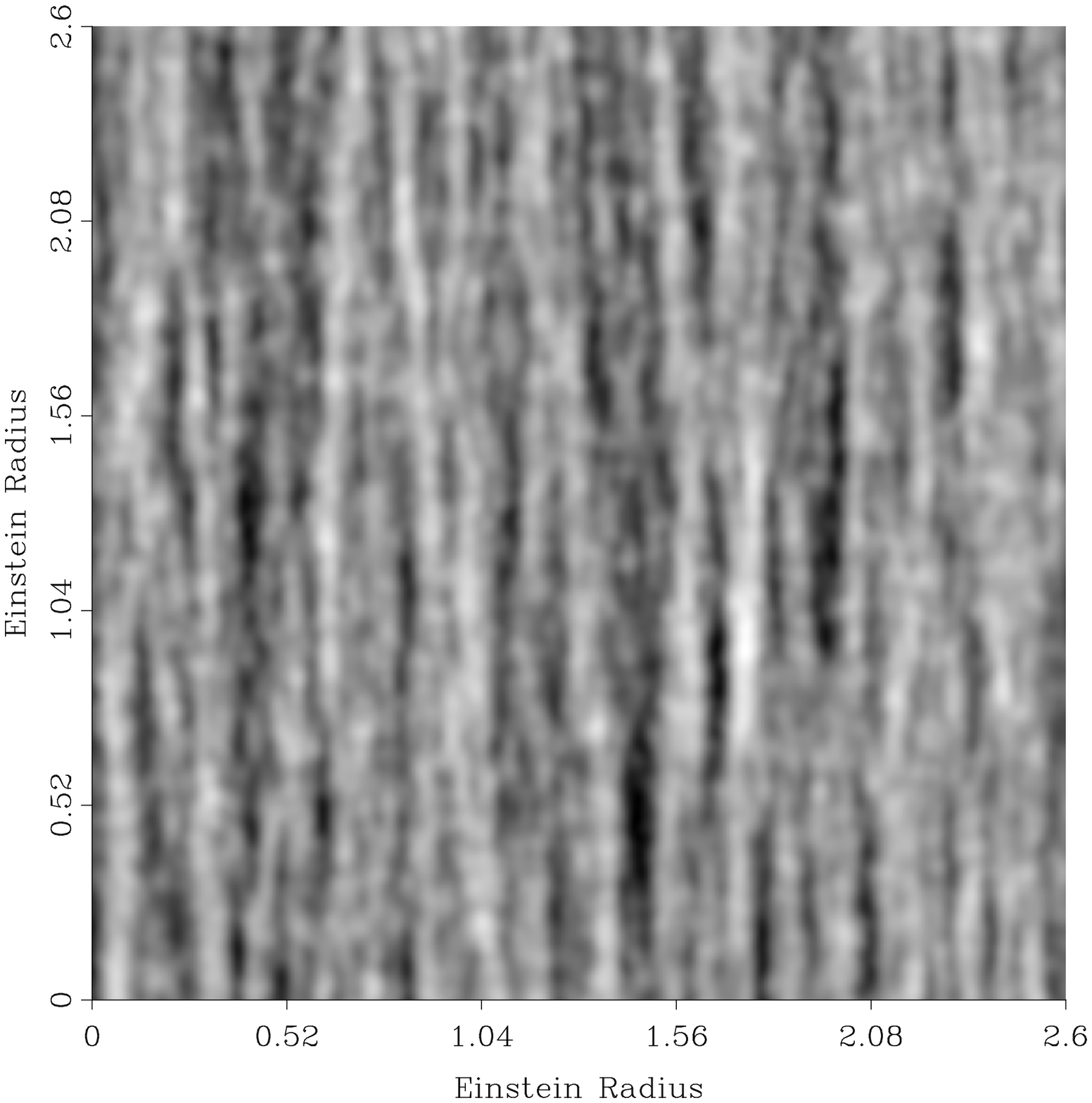}} 
\hspace{2mm}
\subfigure
{\includegraphics[scale=0.48]{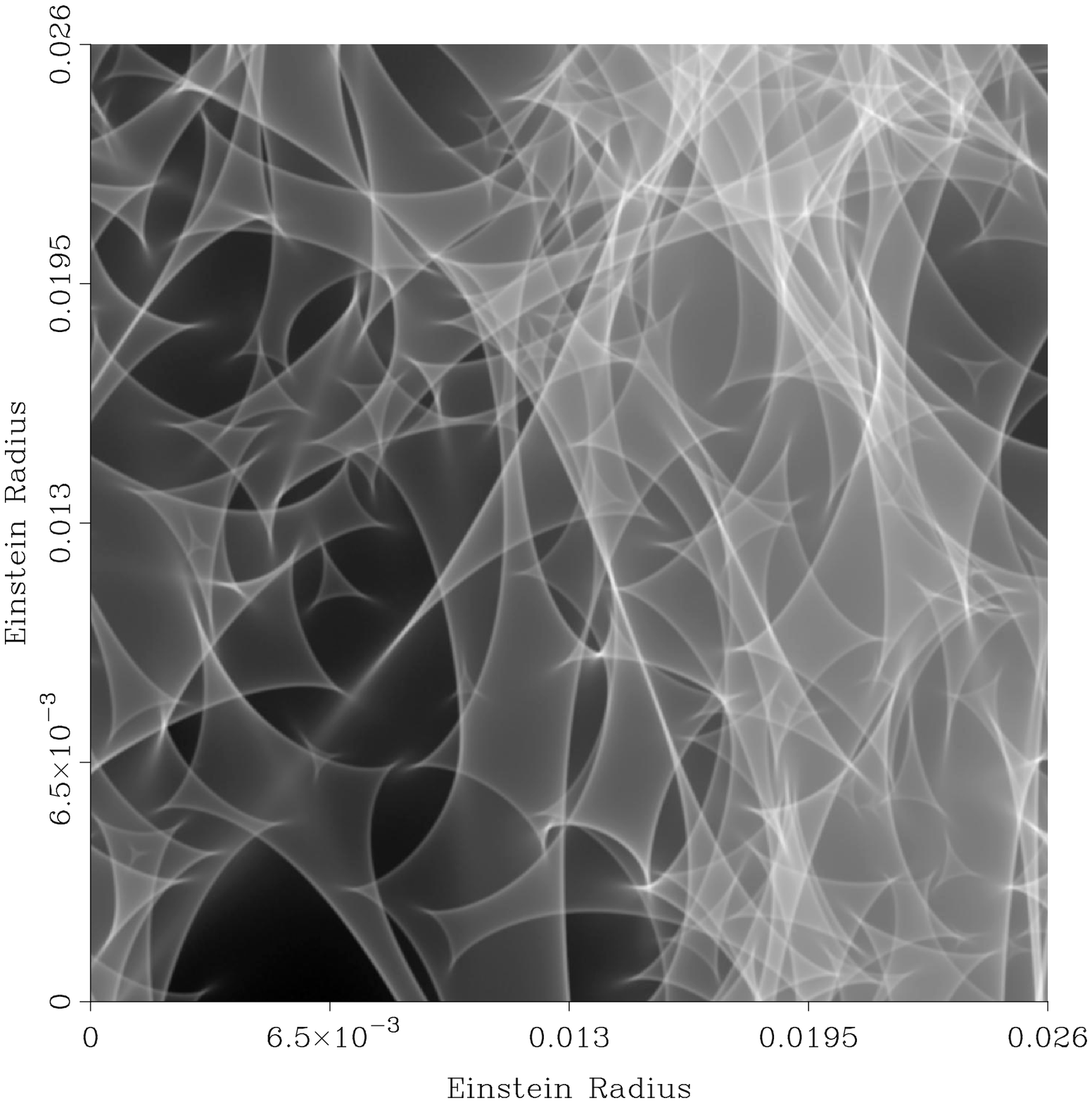}} 
\subfigure{\includegraphics[scale=0.6]{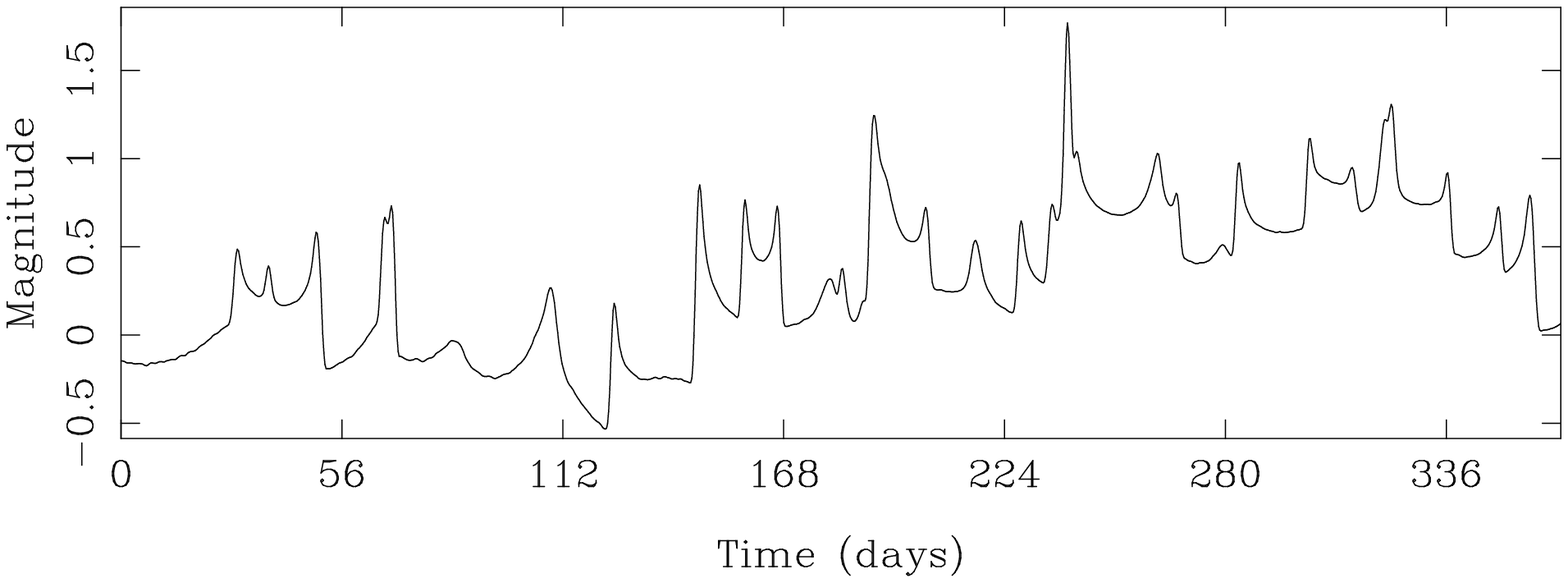}}  
\caption{(a) shows a square region of width 2.6 ER (100 years) cut from the image M map in Figure \ref{maps strip} that was generated with small objects of $2.25\times 10^{-4}$M$_\odot$. The map has been convolved with a source of size 0.05 ER. The region shows caustic bands which produce variability in the
light curve (e.g. Figures \ref{wiggles_by_mass} and \ref{wiggles_by_size}) due to the presence of the small objects, but do not represent nanolensing caustics, which cannot be resolved at these scales. (b) shows a square region of width 0.026 ER (1 year) cut from a map of width 0.2 ER  that was generated with small objects of $2.5\times 10^{-5}$M$_\odot$. The map has been convolved with a source of size 0.0001 ER. The region shows individual nanolensing caustics. (c) contains the light curve cut horizontally through the region in (b).}
\label{bands}
\end{figure*}

It appears, by looking,  that the nanolensing  variability is easier to distinguish from the large scale (microlensing) variability in image S compared to
image M, because the microlensing variability is greater in S. To confirm this we find and compare the nanolensing and microlensing  variability for image M and S. The nanolensing variability is the difference measure we have been using. The microlensing variability is the range of magnifications within the smooth matter map. We present the results for the smallest mass objects, $2.5\times10^{-5}$M$_\odot$, and a source size of 0.002 ER, the map pixel resolution. The nanolensing variability for these is 31\% of the microlensing variability, in image M, in terms of amplitude. For image S, the value is 16\%; hence it appears easier to separate the nanolensing from the microlensing in image S, at this scale. Naturally this is partly because we have used a bi-modal mass distribution, and not a continuous mass spectrum. We will mention this again, below.

The source size used in Figure \ref{wiggles_by_mass},  0.02 ER, crosses its own radius in  140 days, which is above  the crossing time for the Einstein Radius of a  $2.5\times10^{-5}$M$_\odot$ object (71 days) and therefore  probably above the time-scale of nanolensing caustics.
So what is producing the events shown in Figures \ref{wiggles_by_mass}, and in \ref{wiggles_by_size}? 
To see this we show in Figure  \ref{bands}(a) the square region of 100$\times$100 years around the light curve
for the map with  $2.25\times10^{-4}$M$_\odot$ objects, image M, and a source size of 0.05 ER (0.0012 pc). These are mass and source sizes  that sit in the middle of our ranges. The ``wiggles'' in the light 
curve are caused not by individual nanolensing caustics but by bands of densely packed caustics. The nanolensing caustics are not resolvable
at the sources sizes and time periods of Figures \ref{wiggles_by_size} and \ref{wiggles_by_mass}. To highlight this point, we generate a map of 0.2 ER in width, at the same pixel resolution as the maps in Figure \ref{maps strip}, and use a source of size 0.0001 ER ($2.36\times 10^{-6}$ pc). A square segment of this map,
of width 1 year ($6.1\times 10^{-4}$ pc), is shown in Figure \ref{bands}(b). At this resolution we can see the nanolensing caustics. 
Figure \ref{bands} (c) shows a light curve cut across the map in (b). The highest peak  in the light curve has a width
of about 8 days (5.7$\times 10^{-4}$ ER), however there is a small peak just to the left of it which has a width of only 2 days.
A small source crossing these will produce true nanolensing \emph{events}. The events  are resolvable for this scale of mass with this scale of source size, due  to the fact that the  size of 
0.0001 ER is comparable to the Einstein Radius of the small objects (0.00012 ER) and smaller than
the width of the caustics: 8 days = 0.00057 ER.

The fact that we have so far been looking at caustic bands may explain why the nanolensing variability, compared to the
 microlensing variability,  appears to be smaller in image S compared
 to M. The formation of the caustics bands may  be on a similar scale
 in image M and S, whereas the microlensing caustics  are not -- they
 appear to have a larger magnitude range in image S. However, if
 only true nanolensing and microlensing caustics were compared, we
 may find that the relationship of nanolensing to microlensing comes
 out the same for both images. We leave this for future investigation.

\begin{figure*}
\centering
\includegraphics[scale=0.85]{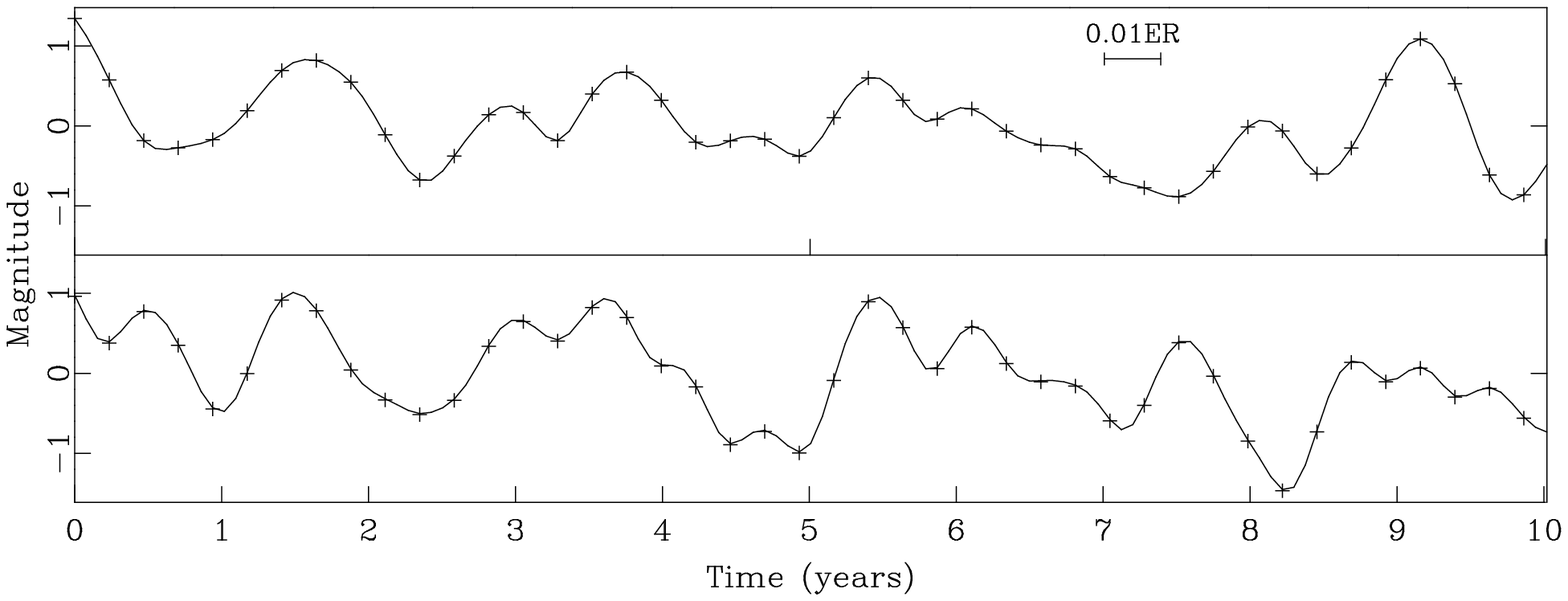}
\caption{Segment of a horizontal light curve taken from the maps in Figure \ref{maps strip} for image M and S, that were generated from small objects of mass $2.5\times 10^{-5}$M$_\odot$ (Figure \ref{maps strip}).  The top row is from the map for image M, the bottom row, image S. The source  size is 0.01 ER.  The segment shown covers a time
period of 10 years. The light curve has been sampled at intervals of 90 days, indicated by the  crosses, showing that
such an interval is sufficient to reconstruct the light curve.
The line labelled 0.01 ER indicates the time period required to cross that distance, i.e. for the source to cross its own radius.}
\label{big_cadence}
\end{figure*}

\begin{figure*}
\centering
\includegraphics[scale=0.9]{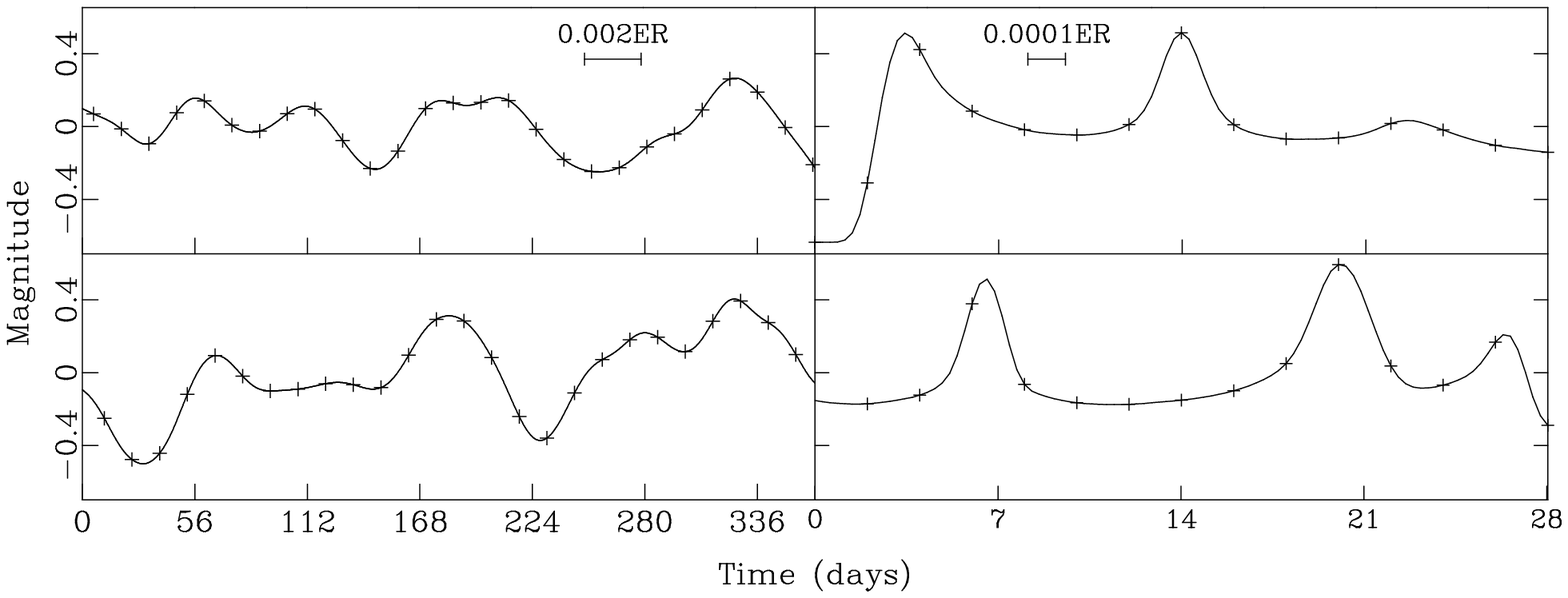}
\caption{Segments of light curves taken from maps generated using  $2.5\times 10^{-5}$M$_\odot$ small objects, with a width of 0.2 ER and a  resolution of
$10000\times10000$ pixel$^2$.
The top row represents image M, the bottom row image S.  The left column is for a source of size 0.002 ER, the right column
for 0.0001 ER. The time period is 1 year for the left column, and 4 weeks for the right column. 
The  crosses indicate samples taken with an interval of 14 days (left column),
or 2 days (right column), showing that observations at these intervals are sufficient to reconstruct the light curve.
The lines labelled 0.002 ER and 0.0001 ER indicate the time period required to cross that distance, i.e. for the source to cross its radius.   The right column shows true (albeit smoothed) nanolensing caustic peaks. The left column does not,  the peaks have been smoothed and combined. }
\label{cadence}
\end{figure*}

We noted in Section \ref{procedure} that when a map generated from stars$+$smooth matter is subtracted from the corresponding map generated from stars$+$small masses, the residual magnitudes have a roughly Gaussian distribution. The residual magnitudes are the nanolensing variability, suggesting that Gaussian noise in microlensing observations will be indistinguishable from  nanolensing. However,  subtraction of the maps in Figure \ref{maps strip} resolves only the caustic bands,  which produce the variability seen   in  Figures \ref{wiggles_by_mass} and \ref{wiggles_by_size}. If variability due to nanolensing caustics  could be resolved,  we  expect the magnitude distribution to assume a shape  more typical of microlensing \citep{lewis1995b,schechter2004}. This would make it possible to distinguish nanolensing from instrument noise. Another method of distinguishing noise be may be to examine the power spectrum of the lightcurves in  Figures \ref{wiggles_by_mass} and \ref{wiggles_by_size}. By inspection,  the caustics bands produce a  periodicity in the lightcurve that should appear as a peak in a power spectrum; such a spectrum will distinguish nanolensing from white and other ``colours'' of noise. A detailed investigation of noise
issues will be set aside for future work.

\subsubsection{Observing cadence}

As seen in Figures \ref{wiggles_by_mass} and \ref{wiggles_by_size}, we find that events have a time scale on the order of a year or less. Figure \ref{big_cadence} shows
a 10-year long segment of the light curve for the case of the $2.5\times 10^{-5}$M$_\odot$ masses, source size  0.01 ER,
for image M (top row) and image S (bottom row); that is, corresponding to the middle row in Figure \ref{wiggles_by_size}. The light curves have been sampled at regular intervals of 90 days, and the resultant points
plotted in crosses on the  light curve. This indicates that an observing cadence of 90 days is enough
to see the nanolensing variability produced by the caustic bands.

To find out how often observations need to be made to observe true nanolensing caustic crossings, we use light curves from the map of width 0.2 ER, created with small objects of $2.5\times 10^{-5}$M$_\odot$. We use two source sizes,
an accretion disk of 0.002 ER, and the smallest source size used previously: 0.0001 ER.
Light curves are cut horizontally across the map and a representative segment
extracted, showing  nanolensing events due to nanolensing caustics. These are in Figure \ref{cadence}, where
the top row is for image M, the bottom for image S; the left column is the source of 0.002 ER, the right column is 0.0001 ER.
The time span is 1 year in the left column, and 4 weeks in the right column. 
The light curves are sampled at 14 days (left column) and 2 days (right column) which we suggest is the longest sampling time that will accurately reconstruct these light curves. 
Only the right column shows true (albeit smoothed) nanolensing caustic crossings -- the left column does not. In the left column, nanolensing caustics have been smoothed and combined. Therefore the source of 0.002 ER is still too large to truly resolve the nanolensing caustics -- a  size of  order 0.0001 ER is needed. 

At 814 nm (271 nm in the rest frame), the predicted flux size of the accretion disk in MG 0414+0534 is  0.002 ER = 2$\times 10^{14}$ cm  \citep{mosquera2011}. We assume that the disk is a Shakura-Sunyaev thin disk
\citep{shakura1973} where the
size of the disk ($R$) is given by 
\begin{equation}
R = R_0 \left(\frac{\lambda}{\lambda_0}\right)^{4/3}
\label{R_eqn}
\end{equation}
where $R_0$ is the radius at $\lambda_0$.

Using this equation and the flux size at 814 nm as a reference, we determine that an accretion disk of size  0.0001 ER in the mock lens system would be emitting at a wavelength of 25 nm, i.e. the extreme UV. At shorter wavelengths such as these the emitting region is likely to be highly variable, and such intrinsic variability could interfere with observations of nanolensing events. 
On the other hand, the time delay between images in MG 0414+0534 is $<$ 1 day \citep{trotter2000}, making it possible to separate the intrinsic variability from the nanolensing events in that quasar, since intrinsic variability will be correlated between the images, and nanolensing will not. We also note, however, that 0.0001 ER is a very small source for a high redshift quasar. \citet{mosquera2011} give a mass of 1.82$\times10^9$M$_\odot$ for the super-massive black hole (SMBH), which means the smallest innermost stable circular orbit (ISCO), using an extremal co-rotating Kerr black hole ($R_{ISCO} = GM/c^2$), is 2.7$\times 10^{14}$ cm = 8.7$\times 10^{-5}$ pc = 0.003 ER for the mock lens. This is  above the source sizes we have used in Figure \ref{cadence}.

If the Einstein Radius were altered to be the same as that in MG 0414+0534, 
the parsec source sizes given previously will scale down by a factor of 0.58. The source velocity in MG 0414+0534 is 270 km s$^{-1}$, and with the different Einstein Radius means the light curve durations scale up by  a factor of 1.27. In Figure \ref{cadence} the durations in the left and right columns become 463 days and 35 days respectively, with the sampling points occurring every 17 days and
2.5 days. The source size for the left column becomes 0.0011 ER and the right column 0.00006 ER. The observing cadence has not altered a great deal, but the source needed to resolve nanolensing caustics is now becoming improbably small.


\section{Conclusion}
\label{conclusions}

We have shown that high-quality maps generated from very large numbers of  compact masses can be analysed with  simple statistical measures. Using a root-mean-square measure, we show that if the source size is
increased the microlensing by compact objects and microlensing by smooth matter become indistinguishable, but below
a threshold it is possible to discriminate between small masses in a lens using the amplitude of nanolensing variability. It will require small sources, of order 0.1  ER, to distinguish very small masses from smooth matter at some reasonable level of detection magnitude, e.g. 0.1 mag.  In future work we will increase the number of data points in Figure \ref{diff_versus_mass}
by generating  maps with other small mass values; this is  computationally expensive compared to fleshing out the range of sources, because map generation is more expensive compared to source convolution. However, with  new tools \citep{garsden2010}, this may be feasible. The method also needs to be applied not only to  MG0414+0534, which will require billions of lens masses, but to other lensed quasars. Rather than using a bi-modal mass function, other distributions  such as Salpeter can be tested, and the percentage of small to large objects \citep{pooley2009} can  be varied.

It will also be necessary to obtain a statistical variance of the results by generating  maps that are identical to those in Figure \ref{maps strip} but with
different randomly-generated locations for the objects. These maps can be passed through the analysis and the results compared to those presented here. A slight difference is expected as the large-scale caustics in the compact matter maps may shift slightly due to a different field of small objects surrounding them. However, these differences should not be significant.

The analysis of the light curves show that there are three scales or levels of lensing in this study -- microlensing due to 
the 1M$_\odot$ stars, variability due to the bands of nanolensing caustics, and peaks due to the nanolensing caustics themselves. Further work needs to be done to determine the scale at which the nanolensing caustics resolve out of the caustic bands, and if, at the different scales, and in different images, the amplitude and duration of nanolensing variability remains the same. It is possible that such analyses will  provide tighter constraints on the masses within the lens, rather than the general analysis here.

Nanolensing events are very small. Even sources as small as the optical accretion disk of MG 0414+0534 will not be enough to resolve individual nanolensing caustics accurately, and smaller  regions will be necessary to do this. Such
regions are less than the expected ISCO of the SMBH, but if they exist, and are not too intrinsically variable,    an observing cadence of 2 days will resolve nanolensing caustics.  An observing proposal could be made on this basis, either using the HST or
Magellan telescopes, for observations of nanolensing events in a quasar such as MG 0414+0534.


\section*{Acknowledgments}

Computing facilities were provided by the High Performance Computing Facility, University of Sydney, 
and the NCI National Facility at the ANU. Hugh Garsden acknowledges funding support from an Australian Postgraduate Award.
This work is undertaken as part of the
Commonwealth Cosmology Initiative (www.thecci.org), and funded
 by the Australian Research Council  Discovery
 Project DP0665574. We thank  the anonymous reviewer whose comments helped us to improve the quality of the paper.

\bsp

\label{lastpage}

\end{document}